\documentclass[journal,transmag]{IEEEtran}
\usepackage[utf8]{inputenc}
\usepackage{bm}
\usepackage{amsfonts}
\usepackage{amsmath}
\usepackage{graphicx}

\begin{document}

\title{Microstructure Role in Permanent Magnet Eddy Current Losses}

\author{\IEEEauthorblockN{Salvatore Perna\IEEEauthorrefmark{1},
Thomas Schrefl\IEEEauthorrefmark{2},
Claudio Serpico\IEEEauthorrefmark{1}, 
Johann Fischbacher\IEEEauthorrefmark{2}, and
Andrea Del Pizzo\IEEEauthorrefmark{1}}
\IEEEauthorblockA{\IEEEauthorrefmark{1}University of Naples Federico II, Department of Electrical Engineering and Information Technology, Naples 80125, Italy}
\IEEEauthorblockA{\IEEEauthorrefmark{2} Danube University of Krems, Department for Integrated Sensor Systems, Wiener Neustadt, Austria}
\IEEEauthorblockA{}% <-this % stops an unwanted space
\thanks{Corresponding author: S. Perna (email: salvatore.perna@unina.it)}}

% The paper headers
\markboth{GG-08}%
{}

\IEEEtitleabstractindextext{%
\begin{abstract}
The impact of granular microstructure in permanet magnets on eddy current losses is investigated. A numerical homogenization procedure for electrical conductivity is defined. Then, an approximated simple analytical model for the homogenized conductivity able to capture the main features of the geometrical and material dependences is derived. Finally eddy current losses analytical calculations are given, and the two asymptotic expressions for losses in the stationary conduction limit and advanced skin effect limit are derived and discussed. 
\end{abstract}

\begin{IEEEkeywords}
eddy currents, homogenization, losses
\end{IEEEkeywords}}

\maketitle

\section{Introduction}
Permanent magnet synchronous motors (PMSMs) represent the best candidate to improve the technological limits of the induction motor in terms of power density and torque density. This is mainly due to presence of permanent magnets in the rotor where Joule losses are much lower than those in induction motors with squirrel cage or rotor windings. %Therefore, PMSMs allow to reduce motor dimensions for fixed power or increase the power limit for fixed  dimensions satisfying the temperature constraint of the insulation classes\cite{IEC}. 
The main issue related to the adoption of PMSMs is the demagnetization effect in permanent magnets and the consecutive reduction of efficiency \cite{Adly_2019,He_2015}. %Indeed, as a consequence of demagnetization there is a reduction of magnetic air-gap flux and then of torque.
From a physical point of view, demagnetization in permanent magnets is a very complex process which depends on many factors such as: external magnetic field, temperature, microstructure of the permanent magnet, etc \cite{Ruoho_2007,Woodcock_2012}.  %The power losses responsible for the increasing of temperature can be divided in winding losses, iron losses, excess losses, and permanent magnet eddy current losses. This last term is usually neglected during the design phase of the permanent magnet \cite{Hafner_2010}, but for high speed applications it has been found to represent a considerable amount of the total losses \cite{Bettayeb_2010}.
One of the causes of the temperature increase in permanent magnet are eddy current losses. They are usually neglected during the design phase of the permanent magnet \cite{Hafner_2010}, but for high speed applications they have been found to represent a considerable amount of the total losses of the motor\cite{Bettayeb_2010}.

\begin{figure}
    \centering
    \includegraphics[width= 8 cm]{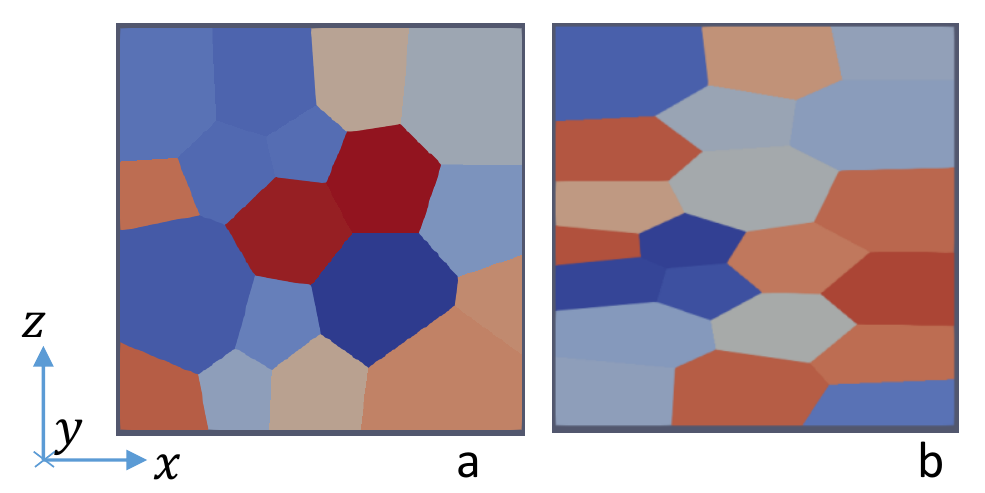}
    \caption{Section in the $xz$ plane of the microstructure for (a) polycrystalline and (b) hot-deformed NdFeB magnets.}
    \label{fig:1}
\end{figure}

In this work, we focus on eddy current losses in permanent magnets and in particular on their dependence on the microstructure of the magnet material (e.g grain dimensions, intergranular phase thickness and electrical conductivity)\cite{Yang_2017}. In the literature, the loss analysis in permanent magnets are based on the solution of the magneto-quasi-static Maxwell’s equations by finite element method (FEM) \cite{Laskaris_2011}, but the cell size of millimeter dimensions does not allow to take into account the microscopic granular structure of the magnet's material. % Therefore, with such an approach, eddy current losses can be studied by changing for example the shape of  magnets  or the electric conductivity assumed to be uniform on the cell-size scale.
In the proposed method the computation of the losses in the whole magnet are based on a two spatial scales approach. On the smaller scale, a micron sized cube is considered and an homogenized electrical conductivity which takes into account the magnet's microstructure is defined. For this quantity, an approximated analytical expression  able to capture its main microstructure dependences is also derived. The granular microstructures considered are shown in fig.\ref{fig:1}. They represent a polycrystalline granular microstructure (see fig.\ref{fig:1}-a) and a hot-deformed granular microstructure (see fig.\ref{fig:1}-b) in sintered NdFeB magnets. The first are universally used in the electric propulsion with PMSMs. However, hot deformed NdFeB magnet are good candidates to substitute polycrystalline sintered magnets in application where high coercivity is required \cite{Li_2018}.
At this point, on the bigger spatial scale, the eddy current losses are computed by solving the eddy current problem in the whole magnet with the homogenized electrical conductivity. %It is shown that the solution of this problem well approximate the solution of the eddy current problem where spatial variations of the electric conductivity due to the microstructure are considered. 
%This permits to connect the eddy current losses computation with a certain homogeneous conductivity to a specific microstructure of the magnet. As an example of this methodology the calculus of eddy current losses  in a parallelepiped shaped magnet excited by sinusoidally time variyng and spatially uniform magnetic field is addressed.
As an example of this methodology the calculus of eddy current losses  in a parallelepiped shaped magnet excited by sinusoidally time varying and spatially uniform magnetic field is addressed.
For this problem the analytical solution in terms of fields and losses is found and the two losses asymptotic expressions in the stationary conduction limit (low frequency) and advanced skin effect limit (high frequency) are derived. 
With such expressions the role of the magnet's homogeneous conductivity and then of the microstructure in eddy current losses is discussed. Finally, design criteria in connection to the application are also discussed.\\
 
\section{Homogenization}
The electromagnetic system under investigation is a uniformly magnetized permanent magnet immersed in a spatially uniform and sinusoidally in time varying magnetic field. The magnet material has a granular microstructure like one of those shown in figure \ref{fig:1}. As a general feature, the microstructure shows the presence of grains separated by a relative thin layer which is the intergranular phase. The grains have uniform electrical conductivity $\sigma$ which differs from that of the integranular phase. From a numerical point of view, to take into account such a variation in a finite element code, requires a discretization mesh with billions of nodes and elements for a body of centimeter sizes, which is unpractical even for the most powerful calculator system. This aspects motivates the study of the homogenization for $\sigma$. The magnet is assumed to be uniformly magnetized with magnetization not perturbed by the external field. The excitation field frequency is in the range $0-100$ kHz which is a typical range in electric motors, due to inverters,  stator slots, etc \cite{Ouamara_2019}. 
The eddy current problem in the magnet, is studied with the magneto-quasi-static (MQS) Maxwell's equations \cite{Haus_Melcher}, which is formulated as generalized diffusion problem in terms of magnetic vector potential and electric potential $(\bm A,\phi)$ \cite{Mayergoyz_1993,Albanese_1990}. This formulation is the starting point to define the homogenization procedure.

Let us consider is a micron sized cube $\Omega_c$ in the permanent magnet. Typical conductivity values and excitation frequency ranges result in a skin depth of $\delta \sim 1/\sqrt{\mu_0\sigma\omega}\sim 10^{-3}\div 10^{-2} $ m, which is several order of magnitude bigger than the lateral size of the cube. This means that the eddy current problem for the cube is reduced to the problem of the stationary conduction where the potential vector can be assumed to be uniform. In this respect, the potentials $\phi$ in $\Omega_c$ satisfies the following equations: 
\begin{equation}\label{eq:eddy_curr_form_2}
\begin{aligned}
&\nabla \cdot \left[\sigma\left(\nabla \phi + \frac{\partial \bm A}{\partial t}\right)\right] = 0\,& in \,\Omega_c\,\\ 
&\hat{\bm n}\cdot [\![\sigma\left(\nabla \phi + \frac{\partial \bm A}{\partial t}\right)]\!] = 0\,&on\,\partial\Omega_c^{int}\,\cup\,\partial\Omega_c\,,
\end{aligned}
\end{equation}
where $\partial\Omega_c$ is the boundary of the cube and $\partial\Omega_c^{int}$ is the internal interface between grain and intergranular phase.
It is immediate to see that the solution of the so formulated problem is not independent from the solution of the eddy current problem defined on the whole magnet. For this reason appropriate boundary conditions able to uncouple the two problems have to be found. In this respect, it is useful to decompose $\phi$ according to the following expression:
\begin{equation}\label{eq:phi_decomp}
    \phi = \sum_{i=1}^3 a_i\,\phi_i\,,
\end{equation}
where $a_i = \frac{\partial\bm A}{\partial t}\cdot \hat{\bm e}_i$ and $\hat{\bm e}_i$ the unit vector of the Cartesian framework. Due to the linearity of the problem \eqref{eq:eddy_curr_form_2}, the potentials $\phi_i$ satisfy the following equations:
\begin{equation}\label{eq:_phi_i_1}
\begin{aligned}
 &\nabla \cdot \left(\sigma\left(\nabla \phi_i + \hat{\bm e}_i\right)\right) = 0\,& in\,\Omega_c\,,\\  
 &\hat{\bm n} \cdot [\![\sigma\left(\nabla \phi_i + \hat{\bm e}_i\right)]\!] = 0\,& on \,\partial\Omega_c^{int}\,.\\
\end{aligned}
\end{equation}
In this respect, the generic potential $\phi_i$ is due to the unit vector $\hat{\bm e}_i$ and equation \eqref{eq:phi_decomp} expresses a superposition principle.
Since the conductivity $\sigma$ oscillates between the conductivity value of the grain and the value of the intergranular phase and it is discontinuous at the interface between them, the normal component of $\nabla\phi$ is discontinuous too and changes sign when $\hat{\bm n}\cdot\hat{\bm e}_i \neq 0$. Therefore, the functions $\phi_i$ are oscillating in the cube and it is reasonable to assume periodic boundary conditions, which means:
\begin{equation}\label{eq:_phi_i_2}
\begin{aligned}
 &\phi_i(x_j = -L/2) = \phi_i(x_j=L/2) \,&on\,\partial\Omega_c\,,\\
 &\hat{\bm n}_j\cdot\nabla\phi_i(x_j = -L/2) = -\hat{\bm n}_j\cdot\nabla\phi_i(x_j = L/2)\,&on\,\,\partial\Omega_c\,,
\end{aligned}
\end{equation}
where $\hat{\bm n}_j$ is the normal to the cube surface with $x_j = \pm L/2$, where $L$ is the side length of the cube. 
From the solution of the three problems \eqref{eq:_phi_i_1},\eqref{eq:_phi_i_2}, the homogenized conductivity  tensor components $\underline{\sigma}_{ij}$ are defined by the following relation:
\begin{equation}\label{eq:sigma_ij}
    \underline{\sigma}_{ij} = \frac{1}{|\Omega_c|}\int_{\Omega_c}\sigma\left(\nabla \phi_i + \hat{\bm e}_i\right)\cdot\hat{\bm e}_j\,dV\,,
\end{equation}
which can be also written as:
\begin{equation}\label{eq:sigma_ij_2}
    \underline{\sigma}_{ij} = \frac{1}{|\Omega_c|}\int_{\Omega_c}\sigma\left(\nabla \phi_i + \hat{\bm e}_i\right)\cdot\left(\nabla \phi_j+\hat{\bm e}_j\right)\,dV\,,
\end{equation}
since according to equations \eqref{eq:_phi_i_1} and \eqref{eq:_phi_i_2} we have:
\begin{equation}\label{eq:sigma_ij_3}
 \int_{\Omega_c}\sigma\left(\nabla \phi_i + \hat{\bm e}_i\right)\cdot\nabla\phi_j\, dV = 0\,.
\end{equation}
 We point out that from equation \eqref{eq:sigma_ij_2}  that the homogenized tensor is symmetric $\underline{\sigma}_{ij}=\underline{\sigma}_{ji}$ and its definition is equivalent to the following equation:
\begin{equation}\label{eq:avg_J}
    \underline{\bm J} = \underline{\bm \sigma}\frac{\partial \bm A}{\partial t}   =\frac{1}{|\Omega_c|}\int_{\Omega_c}\sigma\left(\nabla \phi + \frac{\partial \bm A}{\partial t}\right)\,dV\,.
\end{equation}
Moreover, it is possible to prove the following formula:
\begin{equation}\label{eq:losses_cons}
    \int_{\Omega_c}\sigma\left(\nabla\phi + \frac{\partial\bm A}{\partial t} \right)^2\,dV = \frac{\partial\bm A}{\partial t}\cdot\underline{\bm \sigma}\cdot \frac{\partial\bm A}{\partial t}\,|\Omega_c|\,,
\end{equation}
which corresponds to say the homogenized tensor is positive definite.
From the physical point of view, given the vector potential vector uniform in $\Omega_c$, the left hand side of equation 
\eqref{eq:losses_cons} represents the eddy current losses in the cube. The validity of such formula implies the following one:
\begin{equation}\label{eq:losses_cons_int}
\int_{\Omega_m}\sigma\left(\nabla\phi + \frac{\partial\bm A}{\partial t} \right)^2\,dV \approx\int_{\Omega_m}\frac{\partial\bm A}{\partial t}\cdot\underline{\bm \sigma}\cdot \frac{\partial\bm A}{\partial t}\,dV\,,
\end{equation}
which means that knowing the potential vector distribution we can compute the losses in the magnet as it is homogeneous but taking into account the magnet's microstructure. The problem of how to find the vector potential distribution will be studied in the next section. At this stage, let us  prove equation \eqref{eq:losses_cons}. According to the decomposition shown in equation \eqref{eq:phi_decomp}, it can be written as: 
\begin{equation}\label{eq:losses_cons_2}
\begin{aligned}
&\int_{\Omega_c}\sigma\frac{\partial\bm A}{\partial t}\cdot\left(\nabla\phi + \frac{\partial\bm A}{\partial t} \right)\,dV=\\ &= \sum_{i,j}a_ia_j\int_{\Omega_c}\sigma\left(\nabla \phi_i + \hat{\bm e}_i\right)\cdot\hat{\bm e}_j\,dV=\sum_{i,j}a_i\,\underline{\sigma}_{ij}\,a_j |\Omega_c|\,,   
\end{aligned}
\end{equation}
while from equations \eqref{eq:phi_decomp} and \eqref{eq:sigma_ij_3}, it can be derived the following one:
\begin{equation}\label{eq:losses_cons_3}
\int_{\Omega_c}\sigma\nabla\phi\cdot\left(\nabla\phi + \frac{\partial\bm A}{\partial t} \right)\,dV=0\,.  
\end{equation}
At this point, it is immediate to see that the sum side by side of the  equations \eqref{eq:losses_cons_2} and \eqref{eq:losses_cons_3} is equal and then proves equation \eqref{eq:losses_cons}.  
\begin{figure}
    \centering
    \includegraphics[width= 8 cm]{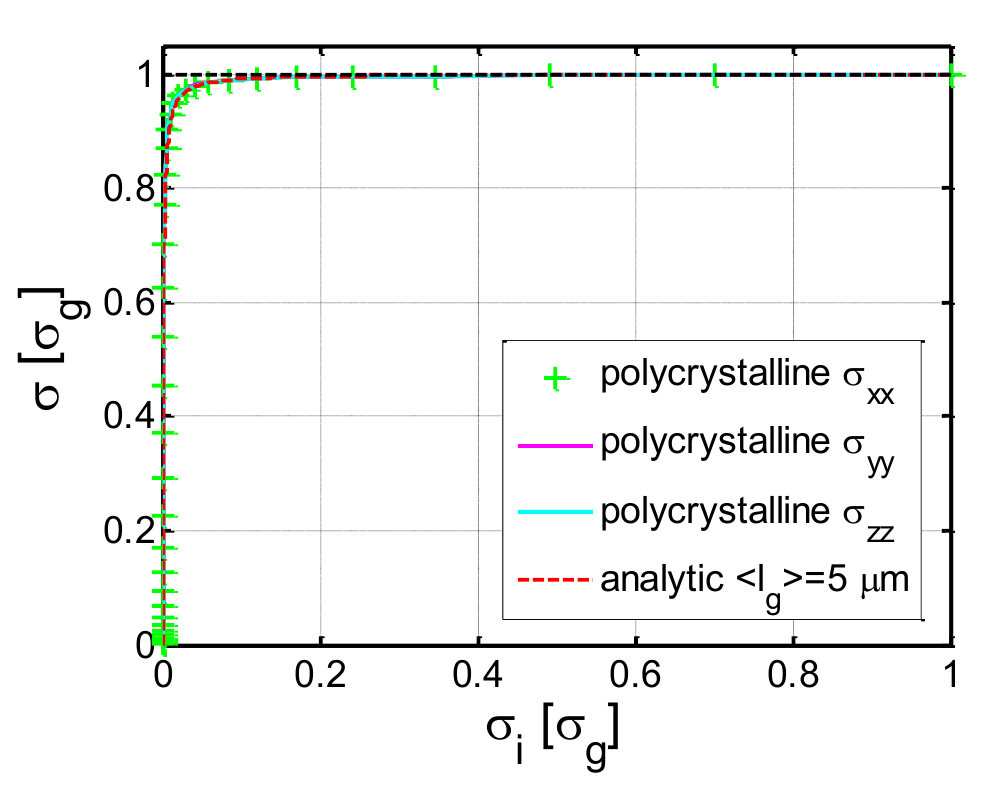}
    \caption{Diagonal component of the homogenized conductivity tensor for a polycrystalline NdFeB magnet in function of the conductivity of the intergranular phase. The number of grains is 1024 and the grains' average side length is $\sim 5\,\mu m$, while the intergranular phase thickness is $5\,nm$}
    \label{fig:2}
\end{figure}
\begin{figure}
    \centering
    \includegraphics[width= 8 cm]{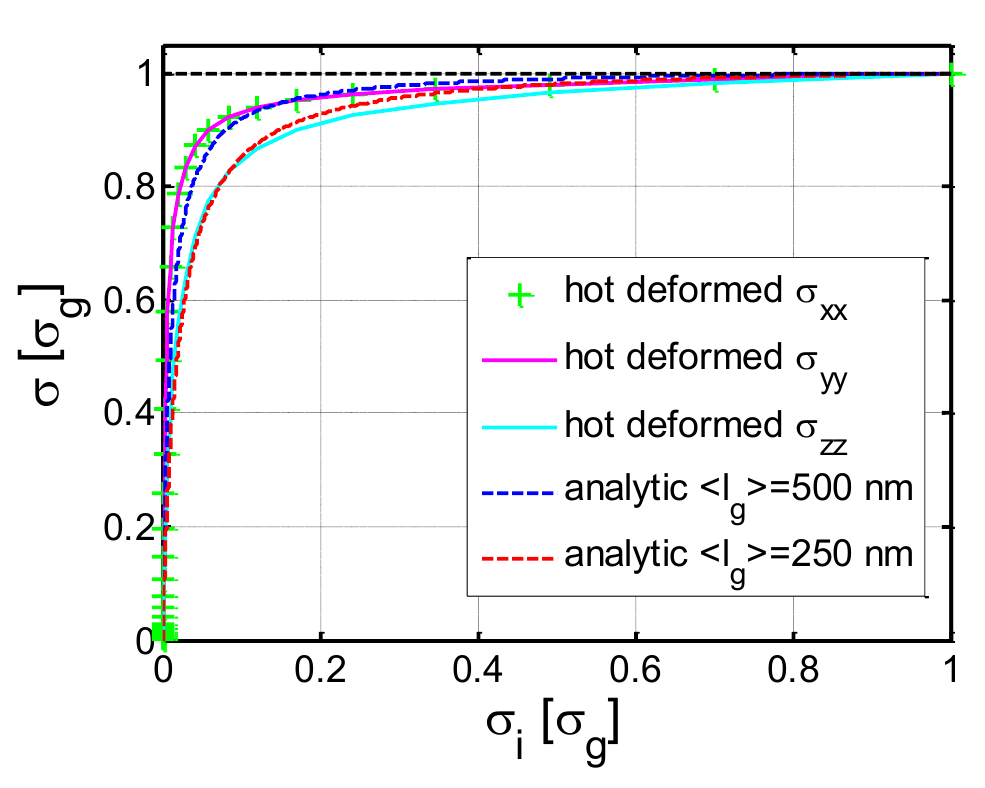}
    \caption{Diagonal component of the homogenized conductivity tensor for a hot deformed NdFeB magnet in function of the conductivity of the intergranular phase. The number of grains is 1024 and the grains' average side length along  the $x$ and $y$ directions is $\sim 500\,nm$ while along the $z$ direction is $\sim 200-400\,nm$. The intergranular phase thickness is $5\,nm$.}
    \label{fig:3}
\end{figure}

\begin{figure}
    \centering
    \includegraphics[width= 8 cm]{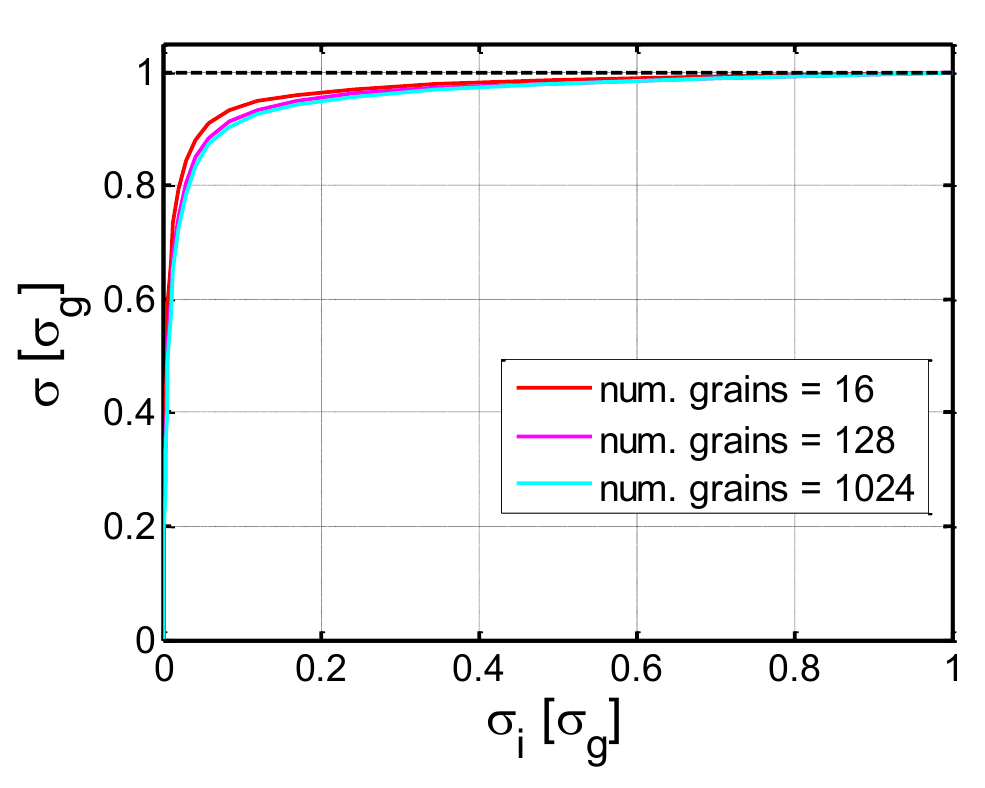}
    \caption{Homogenized conductivity for a polycrystalline NdFeB magnet in function of the conductivity of the intergranular phase and parametrized with the number of grains which constitute the cube microstructure. The grains' average side length is $\sim 500\,nm$, while the intergranular phase thickness is $5\,nm$ }
    \label{fig:4}
\end{figure}
Figures \ref{fig:2} and \ref{fig:3} show the homogenized conductivity tensors calculated from equation \eqref{eq:sigma_ij} for a polycristalline and a hot deformed NdFeB permanent magnets as a function of the conductivity of the intergranular phase.
In the figures $\sigma_g$ and $\sigma_i$ represent the conductivity of the grains and the intergranular phase respectively. The off-diagonal terms of the tensors are found to be negligible for a sufficient number of grains. For polycrystalline magnets it happens that $\underline{\sigma}_{xx}\approx\underline{\sigma}_{yy}\approx\underline{\sigma}_{zz}$, while for the hot deformed magnet we have $\underline{\sigma}_{xx}\approx\underline{\sigma}_{yy}>\underline{\sigma}_{zz}$. Moreover for the two kind of magnets the functional dependence from the integranular phase conductivity changes. Indeed, the hot deformed magnet exhibits a more smooth dependence which produces a lower homogenized conductivity for a wider range of intergranular phase conductivity values.
This dependencies are not altered by the number of grains as it is shown in figure \ref{fig:4} where a single diagonal component of the homogenized conductivity tensor for a polycrystalline magnet is computed increasing the number of grains. Similar results were obtained for the hot deformed magnet.
\begin{figure}
    \centering
    \includegraphics[width= 6 cm]{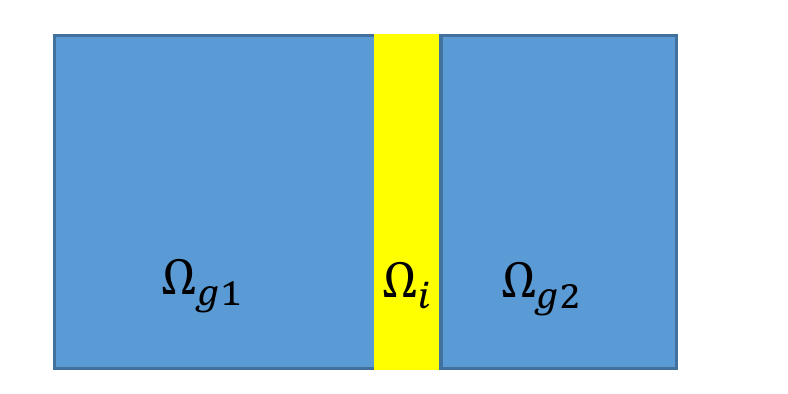}
    \caption{Section in the $xz$ plane of two grains microstructure with parallepiped shape}
    \label{fig:5}
\end{figure}
A simple analytical model which quantitatively describes the dependencies of the homogenized conductivity tensor components from the magnet microstructure can be obtained starting from the two grains microstructure shown in figure \ref{fig:5}. Assuming that $\hat{\bm e}_i = \hat{\bm x}$, the solution of equations \eqref{eq:_phi_i_1}, \eqref{eq:_phi_i_2} and \eqref{eq:sigma_ij}  is reduced to the solution of the following system of algebraic equations:
\begin{equation}\label{eq:_phi_i_3}
\begin{aligned}
 &\sigma_g(f_{xg}+1) = \underline{\sigma}_{xx}& in \,\,\Omega_{g1}\cup\Omega_{g2}\\\  
 &\sigma_i(f_{xi}+1) = \underline{\sigma}_{xx}& in \,\,\Omega_{i}\\
 &2l_g\,f_{xg}+l_i\,f_{xi}=0\,,
\end{aligned}
\end{equation}
where $f_{xg} = \nabla\phi_x\cdot\hat{\bm x}$ in the grains region while $f_{xi}$ is the same quantity but evaluated in the intergranular phase.  The solution of such a system is very straightforward and we find the following expression for the homogenized conductivity $\underline{\sigma}_{xx}$:
\begin{equation}
  \underline{\sigma}_{xx}=\sigma_g\,\frac{\left(1+\frac{l_i}{l_{g1}+l_{g2}}\right)}{\left(1+\frac{\sigma_g}{\sigma_i}\frac{l_i}{l_{g1}+l_{g2}}\right)}\,.   
\end{equation}
The generalization of this formula to the case of a large number of grains $N\gg 1$ with different length $l_{g}$ along the $x$ direction and different intergranular phase length $l_i$ for each couple of grains, can be written as follows:
\begin{equation}\label{eq:approx_homog_sigma}
  \underline{\sigma}_{xx}=\sigma_g\,\frac{\left(1+\frac{(N-1)<l_i>}{N<l_g>}\right)}{\left(1+\frac{\sigma_g}{\sigma_i}\frac{(N-1)<l_i>}{N<l_g>}\right)}\approx\sigma_g\,\frac{\left(1+\frac{<l_i>}{<l_g>}\right)}{\left(1+\frac{\sigma_g}{\sigma_i}\frac{<l_i>}{<l_g>}\right)}\,,   
\end{equation}
where the angular brackets mean the average over the microstructure.
%\begin{figure}
%    \centering
%    \includegraphics[width= 8 cm]{sigma_delta.pdf}
%    \caption{Dependence from the integranular phase thickness of a %diagonal component of the homogenized conductivity tensor for a polycrystalline permanent magnet. Continuous lines are obtained solving numerically the problem \eqref{eq:_phi_i_1}-\eqref{eq:sigma_ij}, while dotted lineas are obtained from the approximated analytical expression in eq.\eqref{eq:approx_homog_sigma}. In the analytical formula the average side dimension of the grains is assumed to be $<l_g>=500\,nm$.}
%    \label{fig:5_2}
%\end{figure}
The comparison of the results obtained from this formula  with the homogenized conductivity tensor components numerically computed on microstructures like those in Figure \ref{fig:1} are shown in Figures \ref{fig:2}, \ref{fig:3}.% and \ref{fig:5_2}.
Setting the parameters $(<l_i>,<l_g>)$ in the order of the intergranular phase thickness and of the average grain size respectively the agreement is quantitative.
It can be noted that this simple model captures also the independence by the number of grains (see fig. \ref{fig:4}). In accordance to this result, when the ratio $<l_i>/<l_g>$ is not so small, it is possible to modulate the conductivity of the magnet by changing the conductivity of the integranular phase. This aspects goes along to the necessity to have smaller grains in order to increase the coercive field of the magnet. Indeed with bigger grains the number of superficial defects due to the fabrication process increases and so does the magnetization switching due to the nucleation of non uniform magnetization configuration which reduced the coercive field of the grain with respect to the magnetization switching by coherent rotation \cite{Brown_Micromagnetics}. In addition, with increasing grain size the local demagnetizing field increases with respect to the exchange energy, causing a reduction of the nucleation field \cite{Bance_2014}.

\section{Eddy current losses}
Equation \eqref{eq:losses_cons_int} shows that once the field $\bm A$ is known, the losses computed with the homogenized conductivity are equivalent to those computed with the not uniform one. However, an equation to compute $\bm A$ in the whole magnet is needed. Due to the fact that the magnetization distribution is uniform, the vector potential $\bm A$ and the induction field $ \bm B$ appreciably change on a length scale (order of skin depth) which is much bigger than the cube dimensions. For this reason, the following equation approximately holds:
\begin{equation}
    \frac{1}{|\Omega_c|}\int_{\Omega_c}\nabla\times\nabla\times\bm A\,dV\approx\nabla\times\nabla\times\bm A\,.
\end{equation}
At this point if we take the diffusion equation satisfied by the potentials $(\phi,\bm A)$ \cite{Albanese_1990} and average on the cube volume, we obtain the following one:
\begin{equation}
 \nabla\times\nabla\times\bm A \approx -\frac{\mu_0}{|\Omega_c|}\int_{\Omega_c}\sigma\left(\nabla\phi + \frac{\partial\bm A}{\partial t} \right)\,dV = -\mu_0\underline{\bm \sigma}\cdot\frac{\partial\bm A}{\partial t}\,,
\end{equation}
which is the equation for the eddy current problem in the magnet with homogenized electrical conductivity. The second equality is based on  equation \eqref{eq:avg_J}. With the above equation, the procedure to compute eddy current losses in a permanent magnet taking into account the granular microstructure of the magnet material is complete.
As an example of its application, in the following the eddy current losses in a parallelepiped permanent magnet excited by a spatially uniform and  sinusoidal time varying magnetic field are considered. 
For such a geometry the eddy current problem can be analytically solved using the 2 dimensional approximation, namely $A_z =0,\,A_x = A_x(x,y,t),\,A_y = A_y(x,y,t)$, where $z$ is the direction of the exciting field.
This situation corresponds to neglect the end-effect due to the finite dimension of the magnet along $z$.  In this respect, from the analytical solution it is possible to derive the following formula power losses by using equation \eqref{eq:losses_cons_int}:
\begin{equation}\label{eq:losses_3}
P = W_0\,\omega\left[\sum_{k = 1}^\infty\frac{\Pi_{xk}}{\gamma_k^2\sqrt{\left(\frac{\gamma_k\delta_x}{\sqrt{2} L_x}\right)^4+1}}+\sum_{k = 1}^\infty\frac{\Pi_{yk}}{\gamma_k^2\sqrt{\left(\frac{\gamma_k\delta_y}{\sqrt{2}L_y}\right)^4+1}}\right]\,,
\end{equation}
where:
\begin{equation}
    \begin{aligned}
    &W_0 = \frac{4\,B_0^2\,L_xL_yL_z}{\mu_0}\,,\,\,\,\gamma_k = (2k-1)\pi\\
    & \Pi_{xk} = \frac{\lambda_{xki}\sinh{(\lambda_{xkr}L_y)}-\lambda_{xkr}\sin{(\lambda_{xki}L_y)}}{\lambda_{xkr}\lambda_{xki}L_y(\cosh{\lambda_{xkr}L_y}+\cos{\lambda_{xki}L_y})}\,,\\
    &\lambda_{xk} = \frac{\sqrt{2}}{\delta_x} \sqrt{\left(\frac{\gamma_k\delta_x}{\sqrt{2} L_x}\right)^2+j}\,,\,\,\,\delta_x = \sqrt{\frac{2}{\mu_0\,\omega\,\underline{\sigma}_{xx}}}\,,\\
    \end{aligned}
\end{equation}
with $\lambda_{xkr} = \Re(\lambda_{xk})$, $\lambda_{xki} = \Im(\lambda_{xk})$. The quantities with subscirpt $y$ can be obtained by changing the subscripts $x\leftrightarrow y$.
In order to understand the role of the homogenized conductivity in the losses when the frequency is changed, it is relevant to consider the low frequency $\delta_i\gg L_i$ and high frequency $L_i\gg\delta_i$ approximations of the expression above. From the low frequency approximation we obtain the following relation:
\begin{equation}\label{eq:losses_approx_1}
    P\approx\frac{4\,L_z}{\mu_0\,\pi^5}B_0^2\,\omega^2\left[\underline{\sigma}_{xx}L_x^4\,\Pi_x+\underline{\sigma}_{yy}L_y^4\,\Pi_y\right]\,,
\end{equation}
where: 
\begin{equation}
    \begin{aligned}
    & \Pi_{x} = \frac{\sinh{\left(\pi\frac{L_y}{L_x}\right)}-\pi\frac{L_y}{L_x}}{\cosh{\left(\pi\frac{L_y}{L_x}\right)}+1}\,,
    & \Pi_{y} = \frac{\sinh{\left(\pi\frac{L_x}{L_y}\right)}-\pi\frac{L_x}{L_y}}{\cosh{\left(\pi\frac{L_x}{L_y}\right)}+1} \,.\\
    \end{aligned}
\end{equation}
\begin{figure}
    \centering
    \includegraphics[width= 8 cm]{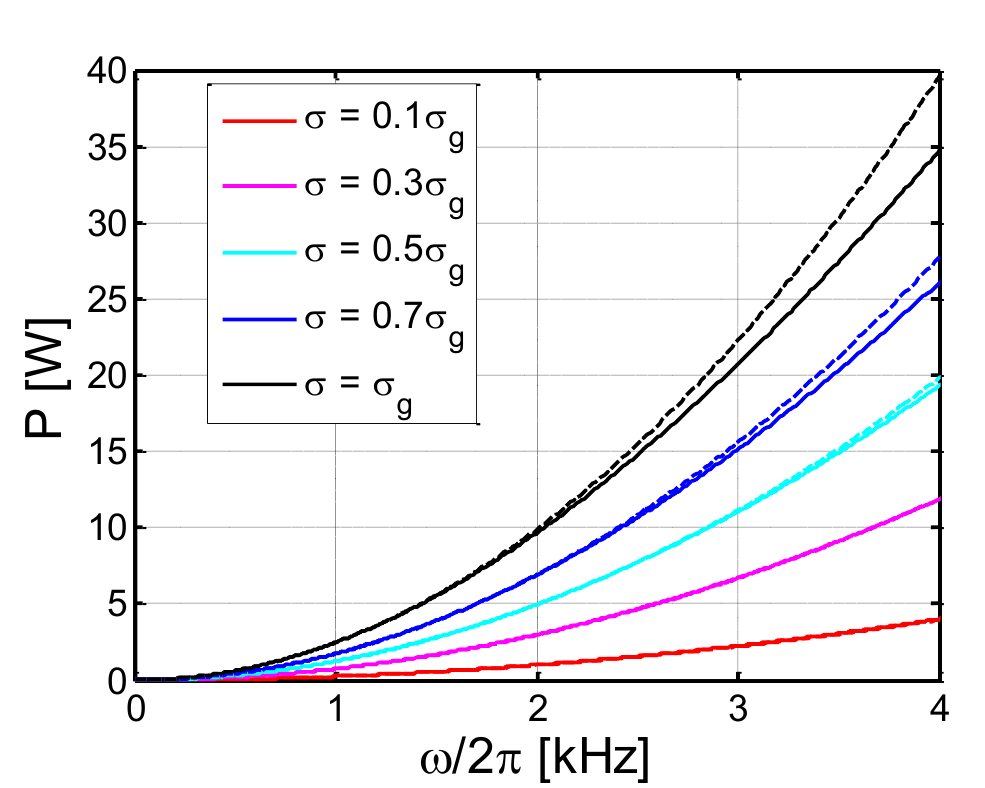}
    \caption{Eddy current losses in the NdFeB permanent magnet in function of the frequency and parametrized with the homogenized conductivity. Continuous lines represent losses computed according to equation \eqref{eq:losses_3}, while dotted lines  represent losses computed according to equation \eqref{eq:losses_approx_1} when $\underline{\sigma}_{xx}=\underline{\sigma}_{yy} = \sigma$.}
    \label{fig:6}
\end{figure}
In figure \ref{fig:6} are shown the eddy current losses for a parallelepiped shaped permanent magnet  with dimensions $L_x = 25\,mm$, $L_y = 14\,mm$ and $L_z = 4\,mm$, which has been used for measurements in reference \cite{Gerlach_2016}. The eddy current losses are plotted in function of the frequency of the exciting magnetic field and are parametrized to the conductivity of the material.
In the figure the losses computed from the formula above and those obtained from equation \eqref{eq:losses_3} in the frequency range $[0,4]\,kHz$ are compared. The agreement is quantitative and improve when the conductivity is reduced. For permanent magnets working in these conditions, to optimize the microstructure means to reduce the homogenized conductivity. This means according to equation \eqref{eq:approx_homog_sigma}, that magnets with small grains, large intergranular thickness and small conductivity of the intergranular phase have better performances in terms of eddy current losses.
For higher frequency, the skin effect is not negligible anymore. When such effect is dominant $ L_i\gg \delta_i$, it is possible to derive an other approximated expression of losses, which is expressed by the following equation:
\begin{equation}\label{eq:losses_approx_2}
    P\approx\frac{5\sqrt{2}\,L_z}{\mu_0^{3/2}\,\pi^2} B_0^2\,\sqrt{\omega}\left(\frac{Lx}{\sqrt{\underline{\sigma}_{xx}}}+
    \frac{Ly}{\sqrt{\underline{\sigma}_{yy}}}\right)\,.
\end{equation}

\begin{figure}
    \centering
    \includegraphics[width= 8 cm]{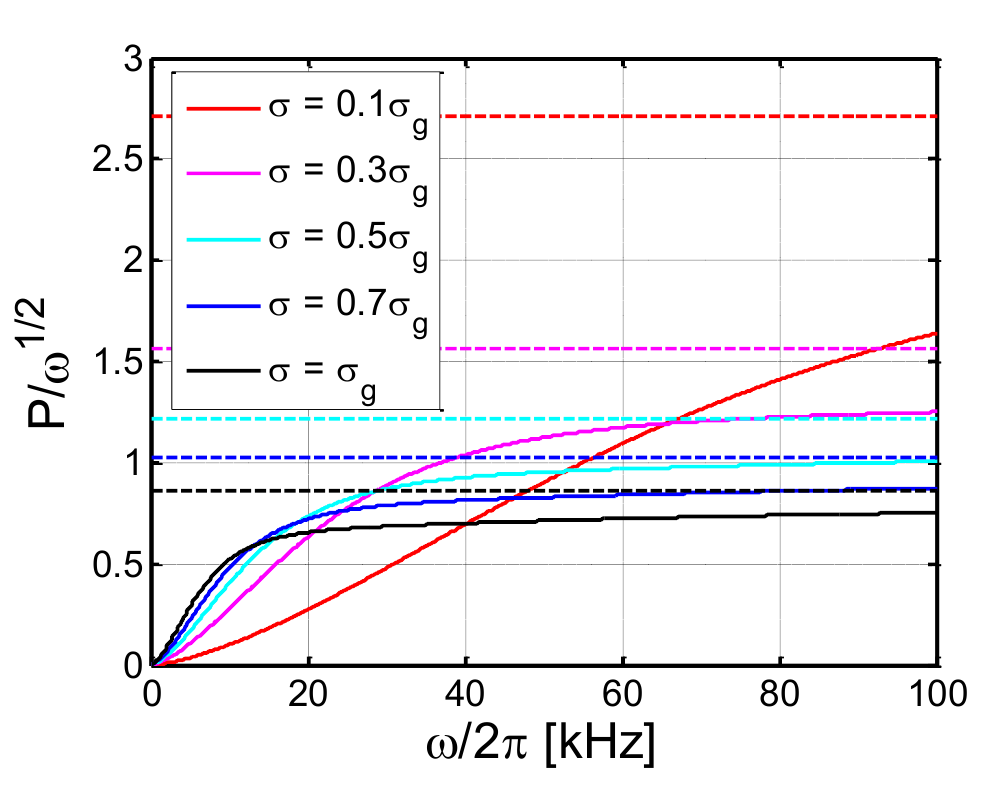}
    \caption{Eddy current losses in the NdFeB permanent magnet in function of the frequency and parametrized with the homogenized conductivity. Continuous lines represent losses computed according to equation \eqref{eq:losses_3}, while dotted lines  represent losses computed according to equation \eqref{eq:losses_approx_2} when $\underline{\sigma}_{xx}=\underline{\sigma}_{yy} = \sigma$.}
    \label{fig:8}
\end{figure}
%This relation is nice for design purpose as equation \eqref{eq:losses_approx_1}. It describes a completely different situation with respect to the previous case where a reduction of conductivity is proportional to a reduction of losses. Here, for certain frequency range, the higher the conductivity the lower the eddy current losses. 
In figure \ref{fig:8} are shown the eddy current losses normalized to the square root of frequency in function of the frequency in the range $[0,100]\,kHz$. From equation \eqref{eq:losses_approx_2} the power loss is plotted by straight dotted lines parallel to the frequency axis. As expected the approximated expression gets more accurate for fixed conductivity as the frequency increases%The curve plotted with straight line are exact losses from equation \eqref{eq:losses_3}, which gets closer to the dotted lines for high value of sigma and of frequency as expected from the assumption made for the derivation of equation \eqref{eq:losses_approx_2}. 
This means that for high frequency, the bigger the conductivity the smaller the losses. This fact is expected, since in the limit of perfect conductor $(\sigma\rightarrow\infty)$ the losses are vanishing.

\section{Conclusions}
In this work the role of granular microstructure of permanent magnets in eddy current losses has been investigated. It is defined an homogenization procedure at the spatial scale of the granular microstructure which under certain assumptions permits to compute the eddy current losses in the whole magnet once the potential vector distribution is known. For the potential vector, an eddy current diffusion equation on the spatial scale of the magnet dimensions is derived, where appears as conductivity coefficient the homogenized electrical conductivity previously defined. Finally as an application of the presented methodology, the eddy current losses are computed for a parallelepiped shaped magnet excited by a spatially uniform and sinusoidally in time varied magnetic field. It is derived that in the low frequency range magnets with small homogenized conductivity, namely small grains and low intergranular phase conductivity, have a better performance in terms of eddy current losses. The same happens in the high frequency range but for magnets with high homogenized conductivity, namely small grains and high intergranular phase conductivity or big grains when for manufacturing reasons the intergranular phase conductivity is low.

\end{document}